\documentclass[twocolumn]{webofc}
\usepackage[varg]{txfonts}   

\begin{document}

\title{Solitonic excitations in collisions of superfluid nuclei}
\subtitle{a qualitatively new phenomenon distinct from the Josephson effect}

\author{
\firstname{Kazuyuki} \lastname{Sekizawa}\inst{1}\fnsep\thanks{\email{sekizawa@if.pw.edu.pl}} \and
\firstname{Gabriel} \lastname{Wlaz{\l}owski}\inst{1,2}\fnsep\thanks{\email{gabrielw@if.pw.edu.pl}} \and
\firstname{Piotr} \lastname{Magierski}\inst{1,2}\fnsep\thanks{\email{piotrm@if.pw.edu.pl}}
}

\institute{
Faculty of Physics, Warsaw University of Technology, ulica Koszykowa 75, 00-662 Warsaw, Poland
\and
Department of Physics, University of Washington, Seattle, Washington 98195-1560, USA
}

\abstract{
Recently, we have reported a novel role of pairing in low-energy heavy ion reactions at energies
above the Coulomb barrier, which may have a detectable impact on reaction outcomes, such as
the kinetic energy of fragments and the fusion cross section [arXiv:1611.10261, arXiv:1702.00069].
The phenomenon mimics the one studied experimentally with ultracold atomic gases, where
two clouds of fermionic superfluids with different phases of the pairing fields are forced to merge,
inducing various excitation modes of the pairing field. Although it originates from the phase difference
of the pairing fields, the physics behind it is markedly different from the so-called Josephson effect.
In this short contribution, we will briefly outline the results discussed in our recent papers and explain
relations with the field of ultracold atomic gases.
}

\maketitle

\section{Introduction}

Although, almost 60 years have passed after the seminal work by A.~Bohr, B.R.~Mottelson,
and D.~Pines \cite{BMP(1958)} that pointed out the existence of the pairing correlations
in atomic nuclei, the pairing dynamics in low-energy heavy ion reactions still remains as one of
the open problems in nuclear physics. On the analogy of the so-called Josephson effect
occuring between superconducting metals \cite{Josephson}, researchers envisaged analogous effects,
referred to as nuclear Josephson effect, leading to subbarrier nucleon-transfer processes in collisions of
two superfluid nuclei \cite{Dietrich1,Dietrich2,Dietrich3,Sorensen}. Nowadays simulations of
heavy ion reactions based on microscopic time-dependent mean-field theories including pairing
correlations have become possible, providing valuable insights into the problem \cite{Scamps(2012),
Scamps(2013),Ebata(2014),Ebata(2015)}, owing to the continuous increase of computation power.
We should note, however, that numerical simulations of nuclear reactions based on full time-dependent
Hartree-Fock-Bogoliubov (TDHFB) theory have been achieved, only very recently \cite{Hashimoto(2016)}.
It can be expected that now the time is ripe enough for elucidation of qualitatively new phenomena
associated with dynamical effects of pairing in low-energy heavy ion reactions.

Due to collaborations of nuclear theory groups at the University of Washington and Warsaw University of Technology,
a theoretical framework, which utilizes a local treatment of time-dependent density functional
theory (TDDFT) including superfluidity, has been extensively developed. It has been shown that the
approach, named time-dependent superfluid local density approximation (TDSLDA), is a powerful
tool for accurately describing complex dynamics of strongly-correlated fermionic systems, like ultracold
atomic gases \cite{PRL__2009,Science__2011,LNP__2012,PRL__2012,ARNPS__2013,PRL__2014,PRA__2015}
and nuclear systems \cite{PRC__2011,PRL__2015,PRL__2016,Mag2016,Vortex}. In our recent papers
\cite{MSW(2017),SMW(2017)}, we have extended the application of TDSLDA to low-energy heavy ion
reactions, employing FaNDF$^0$ nuclear functional \cite{Fayans1,Fayans2} without spin-orbit coupling,
and found a qualitatively new phenomenon associated with the pairing field dynamics in collisions of
two superfluid nuclei.

In Refs.~\cite{MSW(2017),SMW(2017)}, we have reported the results of TDSLDA simulations for
$^{44}$Ca+$^{44}$Ca, $^{90}$Zr+$^{90}$Zr, $^{86}$Zr+$^{126}$Sn, and $^{240}$Pu+$^{240}$Pu
reactions at various incident energies (note that neutrons in ``$^{90}$Zr'' are in superfluid phase without
spin-orbit coupling). It has been shown that a soliton-like excitation mode emerges inside the neck
region, in which pairing vanishes, when two superfluid nuclei possessing different phases of the pairing
field collide. The situation resembles the one encountered in experiments with ultracold atomic gases,
where two clouds of superfluid atomic gases are forced to merge, creating a solitonic excitation
(domain wall) which decays through quantum vortices \cite{MIT1,MIT2}. From the results,
it has been shown that the solitonic structure is relatively long-lived object that stays until the
composite system splits. Because of this fact, the solitonic excitation effectively hinders energy
dissipation from translational motion to internal degrees of freedom as well as the neck formation,
resulting in dramatic changes of the reaction dynamics (see Refs.~\cite{MSW(2017),SMW(2017)}
for more details).

\begin{figure}[t]
   \begin{center}
   \includegraphics[width=\columnwidth]{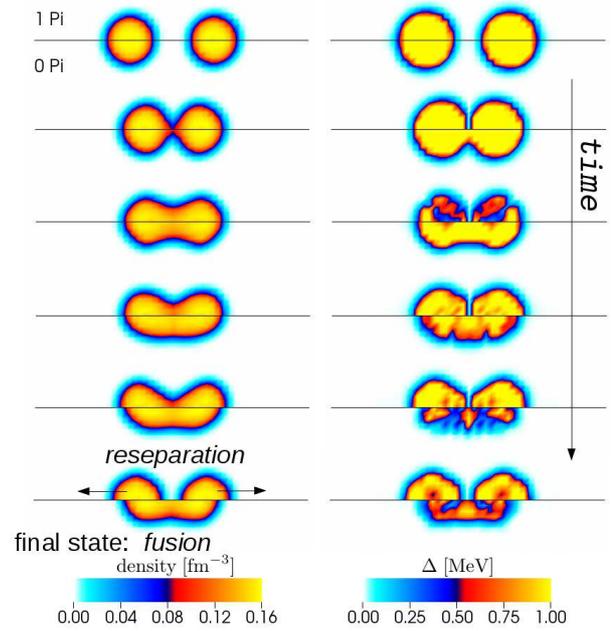}
   \end{center}\vspace{-2mm}
   \caption{
   Snapshots of the total density (left column) and the absolute value of the neutron's pairing field (right column)
   at various times in the TDSLDA simulations for the $^{90}$Zr+$^{90}$Zr reaction at $E\simeq1.05V_{\rm Bass}$,
   where $V_{\rm Bass}$ is the phenomenological fusion barrier \cite{Bass1974}. Upper-half (lower-half) of
   each panel shows those for the $\Delta\varphi=\pi$ (0) case. For full movies see Supplemental Material of Ref.~\cite{MSW(2017)}.
   }\vspace{-3mm}
   \label{fig:Zr+Zr}
\end{figure}

In Fig.~\ref{fig:Zr+Zr}, we show one of the illustrative examples of our TDSLDA simulations for
the $^{90}$Zr+$^{90}$Zr reaction at $E\simeq1.05V_{\rm Bass}$, for two extreme cases:
$\Delta\varphi=\pi$ and $\Delta\varphi=0$, which are shown in upper-half and lower-half of
each panel of the figure, respectively. Here, $\Delta\varphi$ denotes the relative phase difference
of the pairing fields of colliding superfluid nuclei. Snapshots of the total density on the reaction plane
at various times are shown in the left column, while those of the absolute value of the neutron's pairing
field are shown in the right column. Since the collision energy is higher than the Coulomb barrier, two
nuclei collide clearly showing a substantial overlap of the density (see the left column). However,
in the $\Delta\varphi=\pi$ case (upper-half), two nuclei do not fuse and eventually reseparate
generating binary reaction products; whereas the $\Delta\varphi=0$ case (lower-half) resulted
in fusion.

What makes the dynamics so different? The answer lies in the dynamics of excitation modes of the
pairing field, which is depicted in the right column of the figure. From the figure, one can clearly
see that a narrow region in which the pairing field vanishes is created on the course of collision in
the $\Delta\varphi=\pi$ case (upper-half), which we refer to as solitonic excitation of the pairing field.
Since the phase changes so steeply between the colliding nuclei, the system chooses to become
normal inside the solitonic structure. The solitonic excitation prevents the system to form a compact
(mononuclear) shape (see left column) and, as a result, fusion reaction is hindered, even though the
system once overlapped substantially (Note that there is a remarkable difference compared to a lighter
system, where two nuclei easily get fused once the system overlapped only slightly \cite{SMW(2017)}).
Summarizing, from our comprehensive simulations, we have found that \cite{MSW(2017),SMW(2017)}:
total kinetic energy (TKE) of outgoing fragments can be changed as large as 25~MeV in $^{240}$Pu+$^{240}$Pu;
fusion threshold energy can be affected almost 30~MeV in $^{90}$Zr+$^{90}$Zr; similar effects present
also for an asymmetric system, $^{86}$Zr+$^{126}$Sn; whereas the pairing effects turned out to be
small in a lighter system, $^{44}$Ca+$^{44}$Ca, because of the strong tendency towards fusion,
consistent with an earlier study \cite{Hashimoto(2016)}.

Although the effect originates from the relative phase difference of the pairing fields of two
colliding superfluid nuclei, it is not related to the Josephson effect.
As we feel that it would be useful to clarify the distinct character of the effect
we studied, as compared to the Josephson effect, we explain this point in detail.

\section{Solitonic excitation vs Josephson effect}

Let us first remind that the physics of the Josephson effect deals with the phenomenon of
the current flowing through a junction in the absence of the chemical potential difference.
The first Josephson's paper on the subject entitled
{``Possible new effects in superconductive tunnelling''} \cite{Josephson}
states clearly in the introduction that:
{\em ``We here present an approach to the calculation of tunneling currents
between two metals that is sufficiently general to deal with the case when
both metals are superconducting. In that case new effects are predicted,
due to the possibility that electron pairs may tunnel through the barrier
leaving the quasi-particle distribution unchanged''.}

This current, being only dependent on the mutual phase differences of the
pairing fields, is responsible for a variety of effects which are known under
a common name of ``Josephson effect''. However, the crucial point is
that the effect appears from the tunneling through the barrier, and therefore,
the current arises as a consequence of a {\em weak coupling} of the two pair
condensates. Consequently, its magnitude depends on $\sin(\Delta\varphi)$,
where $\Delta\varphi$ is the phase difference between the condensates, as
is dictated by the equations responsible for dynamics of the Josephson effect:
\begin{eqnarray}
J(t) &=& J_{c}\sin[\Delta\varphi (t)], \\[2mm]
\frac{d(\Delta\varphi)}{dt} &=& \frac{2eU}{\hbar},
\end{eqnarray}
where $J_c$ is the amplitude of the Josephson current, $e$ is the elementally charge,
$\hbar$ is the reduced Plank constant, and $U$ is the voltage, which, in nuclear case,
will be proportional to the chemical potential difference. Thus, sub-barrier multinucleon
transfer processes between superfluid nuclei can be viewed as a nuclear Josephson effect
and papers \cite{Dietrich1,Dietrich2,Dietrich3,Sorensen} focused solely on this aspect.

\begin{figure*}[t]
   \begin{center}
   \includegraphics[width=0.5\textwidth]{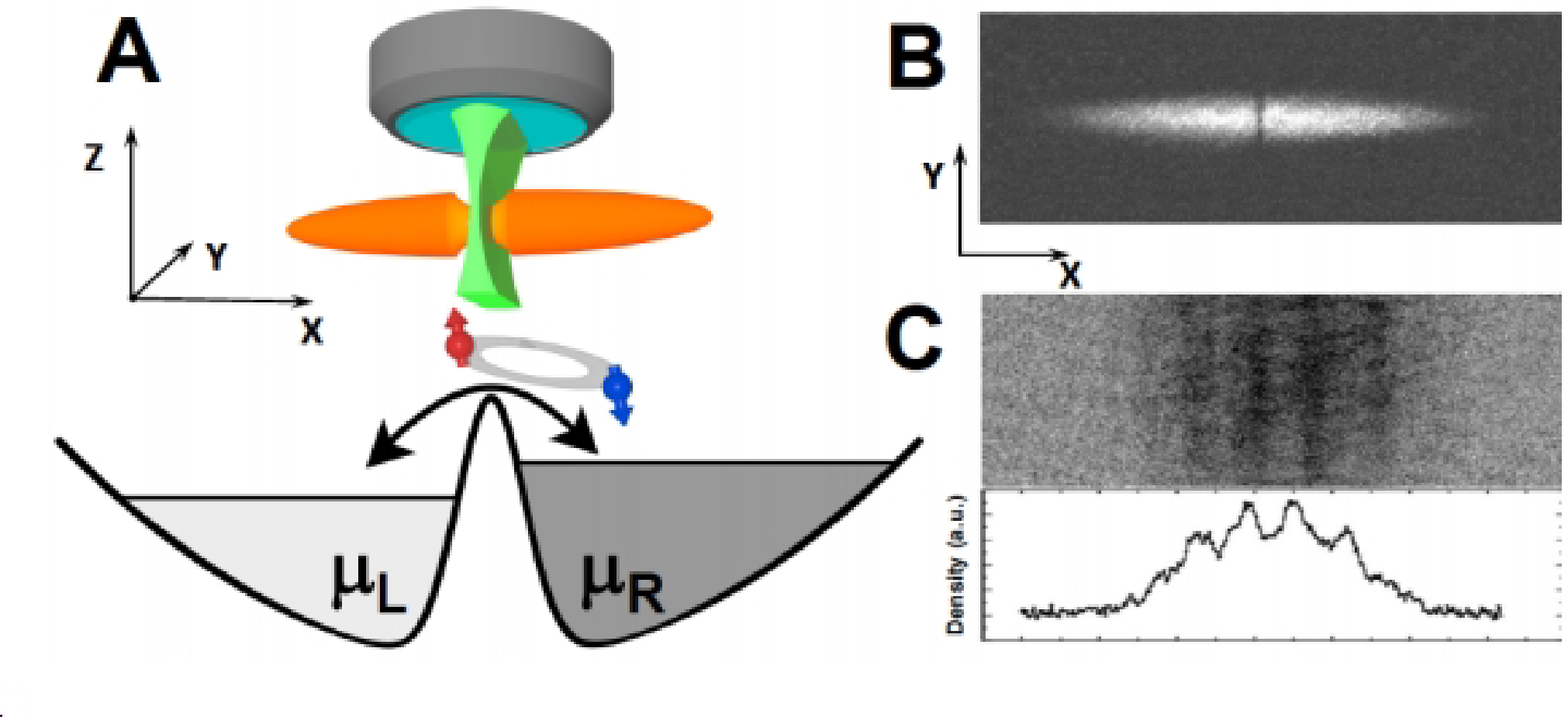}\hspace{3mm}
   \includegraphics[width=0.45\textwidth]{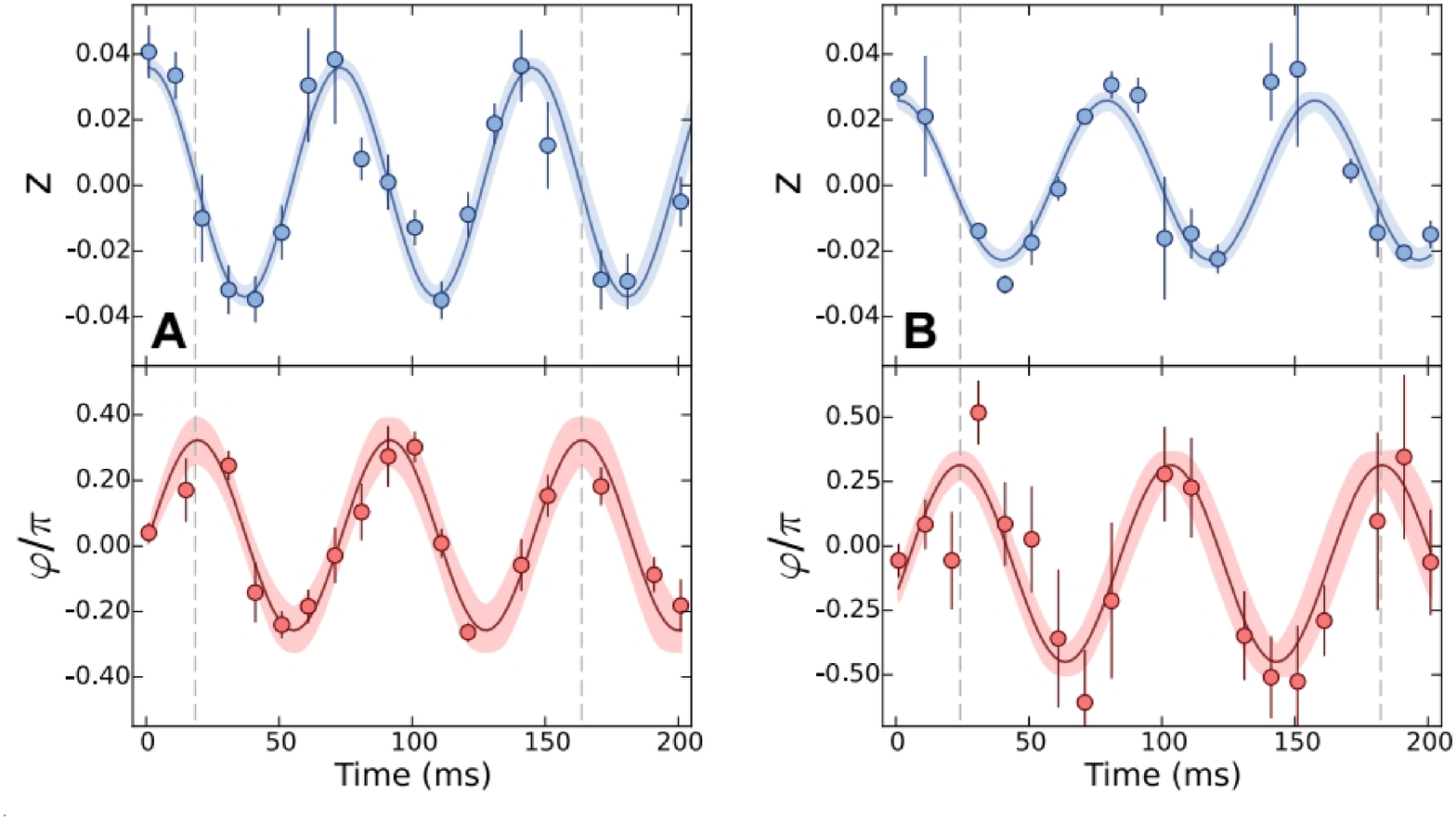}
   \end{center}\vspace{-3mm}
   \caption{
   Left: (A) Sketch of the experimental apparatus for studies of Josephson junction realized
   through introducing potential barrier between two $^{6}$Li condensates, and (B) image of the cloud.
   Right: Oscillation  of the population imbalance $z=\Delta N/N$ and phase difference $\varphi$ between
   fermionic superfluids due to the coherent flow of the particles through the junction. (A) corresponds to
   the BEC regime (B) corresponds to the so-called unitary limit.
   The figure has been reprinted with permission from: G. Valtolina et al., Science {\bf 350}, 1505 (2015) [see this reference for more details].
   }\vspace{2.5mm}
   \label{fig:1}
\end{figure*}

\begin{figure*}[t]
   \begin{center}
   \includegraphics[width=\textwidth]{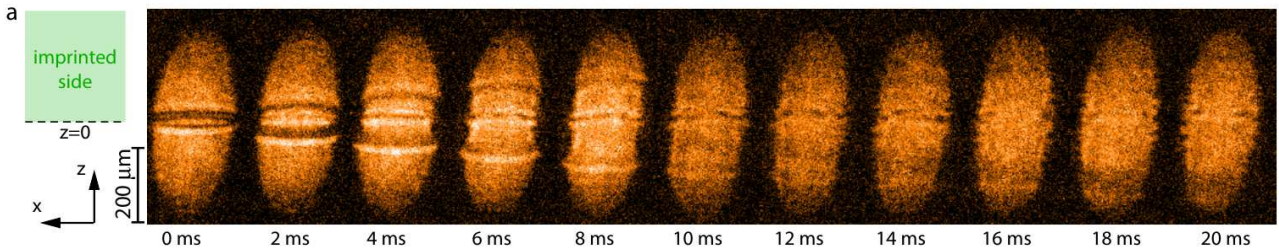}
   \end{center}\vspace{-3mm}
   \caption{
   Evolution of the atomic density when two clouds with different phases are put into contact.
   Just after making contact two sound waves propagating towards the edges are generated.
   In the center dark soliton (visible as darker line) is excited. There is no external barrier that
   separates the cloud. The soliton lives about 10\,ms and decays into a vortex ring (visible as
   two dots on the last panel). The vortex ring is only transient, and is replaced by a vortex line in the final state.
   The figure has been reprinted with permission from: M.J.H. Ku, et al., Phys. Rev. Lett. {\bf 116}, 045304 (2016) 
   [see this reference for more details].}
   \label{fig:2}
\end{figure*}

However, we would like to emphasize here that the Josephson effect described above is
markedly different from the effect we studied in Refs.~\cite{MSW(2017),SMW(2017)}.
What we investigated is collisions of superfluid nuclei at energies above the Coulomb barrier, which,
on the contrary to the Josephson effect, can be regarded as a \textit{strong-coupling} limit,
where two condensates (nuclei) actually merge producing a solitonic excitation (domain wall).
Both effects are associated with the pairing field, but are weakly connected in the sense that:
no solitonic excitation is necessary for the Josephson current to appear; whereas no Josephson
current is necessary for the effect we studied to appear. Indeed, we have shown that the strongest
effect is observed actually for the $\Delta\varphi=\pi$ case where no Josephson current is present.
\cite{MSW(2017),SMW(2017)}. Usually the Josephson effect is studied in the regime of
a quasistationary situation, which occur when two condensates couple weakly through
the potential barrier, and therefore no topological excitation of the pairing field (soliton or
quantum vortex) is expected. In our case, on the other hand, we studied rather violent
collisions at energies above the Coulomb barrier, \textit{i.e.}, far from the stationary situation,
which trigger the appearance of the localized excitation mode between colliding superfluid nuclei.

To emphasize better the differences between the Josephson effect and the effect that
we studied in our papers \cite{MSW(2017),SMW(2017)}, let us consider the system where
both phenomena can be studied with far better accuracy than in nuclear systems. Namely,
let us consider the case of ultracold fermionic (atomic) gases. In the latter case, studies of
the Josephson effect can be achieved experimentally. The current is realized as a tunneling
between two fermionic superfluids separated by an external potential barrier. In Fig.~\ref{fig:1}
(left), we show a typical experimental setup, which allows to investigate a Josephson alternating
current due to the {\em weak coupling} of two Lithium atomic clouds~\cite{valtolina}. As a result
an oscillatory current is observed accompanied with an oscillatory evolution of the phase difference
between the condensates [see Fig.~\ref{fig:1} (right)]. For the flow of the atoms the weak coupling
is sufficient and no solitonic excitation is present.

On the other hand, the effect we studied is more closer to the experiments performed by
the MIT group~\cite{Yesfah,MKu1,MKu2}, which consist in {\em merging} two atomic
clouds with different phases and producing topological excitations (see Fig.~\ref{fig:2}).
There is no external potential that separates the cloud at all. Note that this phenomenon,
although originating from the pairing field dynamics, is not related to the Josephson effect. 
The reason is that in this case physics is different, although of course certain flow of particles
induced by the pairing field is observed. However, the main feature comes from the nonlinear
evolution of the pairing field, which makes the solitonic excitation relatively long lived. 
With TDSLDA approach, we were able to successfully simulate this experiment and we found that 
in case of the unitary Fermi gas the excitation decays into a vortex ring~\cite{Bulgac}, which is
another type of the topological excitations of the pairing field. This prediction was confirmed
by the experiment in Ref.~\cite{MKu2}. 

We understand that in nuclear physics community this distinction has not been emphasized,
since all papers, which concern physics of colliding superfluid nuclei, have investigated only the
consequences of the supercurrent (or pair transfer) which would manifest in multinucleon
transfer processes in heavy ion reactions. Indeed, this effect has been studied, but it is
relatively weak and has nothing to do with the phenomenon we studied in Refs.~\cite{MSW(2017),
SMW(2017)}. The latter requires not only the phase difference but also the creation of the
solitonic structure which stays relatively long so as to hinder energy dissipation and the neck formation
during the collision. In other words, the Josephson effect originates from the weak coupling between
two condensates separated by a potential barrier, whereas in our case we consider a strong coupling
limit where two condensates actually merge/collide. It would be misleading to refer to the latter
as the effect that is already known from the weak coupling limit, in the same way, like it would
be misleading to say that properties of a turbulent flow are known because a laminar flow has
been studied. Analogously, physics of a weakly-interacting Fermi gas described by the standard
Bardeen-Cooper-Schrieffer (BCS) theory is completely different from the one in the strong
coupling limit (\textit{i.e.}~for the unitary Fermi gas) where a variety of qualitatively new
phenomena occur like, \textit{e.g.}, pseudogap or anomalously small ratio of the shear
viscosity to the entropy density. One cannot say that the unitary Fermi gas is a well-known
system because it is a superfluid just like a typical atomic nucleus.

\section{Summary}

In our recent papers \cite{MSW(2017),SMW(2017)}, we have reported a qualitatively new phenomenon
in low-energy heavy ion reactions at energies above the Coulomb barrier, which is associated with relative
phase difference of the pairing fields of colliding superfluid nuclei. We have found that the complex pairing
field dynamics may substantially alter the reaction outcomes, such as the total kinetic energy of the fragments
and the fusion cross section, which may be experimentally detectable for carefully selected systems. In our
simulations, a ``solitonic excitation'' (domain wall) is observed in the neck region, inside which pairing vanishes,
resulting in hindrance of energy dissipation as well as the neck formation, leading to significant changes
of the reaction dynamics. Although this effect originates from the relative phase of the pairing fields of
colliding superfluid nuclei, the physics behind it is markedly different form the so-called Josephson effect:
the latter is associated with a weak coupling of condensates (separated by a potential barrier) without
any solitonic excitation, whereas the effect we studied is associated with a strong-coupling limit where
two condensates are actually merge/collide for which no Josephson current is necessary. The ongoing
further extensions and applications of the present work will enable us to find qualitatively new phenomena
associated with the pairing field dynamics that have never been investigated in nuclear physics community.

\section*{Acknowledgments}

This work was supported by the Polish National Science Center (NCN) under Contracts No.
UMO-2013/08/A/ST3/00708. The code used for generation of initial states was developed under
grant of Polish NCN under Contracts No. UMO-2014/13/D/ST3/01940. Calculations have been
performed at HA-PACS (PACS-VIII) system---resources provided by Interdisciplinary Computational
Science Program in Center for Computational Sciences, University of Tsukuba.

\end{document}